\documentclass{ws-procs961x669}            
\newcommand{\nord}[1]{{:}#1{:}}
\def\luv{\ell_{\rm UV}}

\begin{document}
\title{A singularity theorem for evaporating black holes}

\author{ E.-A. Kontou\footnote{Presenting author}, B. Freivogel}

\address{ITFA and GRAPPA, Universiteit van Amsterdam
\\
	Science Park 904, Amsterdam, the Netherlands\\
E-mail: e.a.kontou@uva.nl}

\author{D. Krommydas}

\address{Instituut-Lorentz, Universiteit Leiden\\
P.O. Box 9506, 2300 RA Leiden, The Netherlands}

\begin{abstract}
The classical singularity theorems of General Relativity rely on energy conditions that are easily violated by quantum fields. Here, we provide motivation for an energy condition obeyed in semiclassical gravity: the smeared null energy condition (SNEC), a proposed bound on the weighted average of the null energy along a finite portion of a null geodesic. Using SNEC as an assumption we proceed to prove a singularity theorem. This theorem extends the Penrose singularity theorem to semiclassical gravity and has interesting applications to evaporating black holes.
\end{abstract}

\keywords{quantum fields, gravity, energy conditions, quantum inequalities, singularities}

\bodymatter

\section{Introduction} 
\label{sec:intro}

The classical or pointwise energy conditions are bounds on components of the stress-energy tensor and were introduced early on in the history of general relativity. Their purpose is to encode properties of what is considered ``reasonable'' matter, and predict the evolution of gravitational systems in a model-independent way. 

The energy condition of interest in this work is the null energy condition (NEC). The NEC states that the stress-energy tensor contracted with two null vectors is non-negative everywhere. Using the Einstein Equation, we get the geometric form of the NEC or the null convergence condition. Using a perfect fluid stress-energy tensor, we can give a physical interpretation of the NEC: the sum of energy density and pressure cannot be negative. The three forms are summarized in Table~\ref{tab:nec}. 

\begin{table}
	\tbl{The different forms of the NEC. Here $\ell^\mu$ is a null vector.}
	{\begin{tabular}{@{}ccc@{}}
			\toprule
			Physical form & Geometric form & Perfect fluid  \\\colrule
			$T_{\mu \nu}\ell^\mu \ell^\nu \geq 0$ & $R_{\mu \nu}\ell^\mu \ell^\nu \geq 0$ & $\rho+P \geq 0$  \\\botrule
	\end{tabular}}
\label{tab:nec}
\end{table}

The NEC is obeyed by minimally coupled scalar fields but as with all pointwise energy conditions, it is violated by quantum fields\cite{Epstein:1965zza}. Ford\cite{Ford:1978qya} was the first to introduce \textit{quantum energy inequalities} (QEIs), restrictions on the possible magnitude and duration of any negative energy densities within a quantum field theory.

QEIs have since been derived for flat and curved spacetimes, bosonic and fermionic fields (see Ref.~\citenum{Kontou:2020bta} and \citenum{Fewster2017QEIs} for recent reviews). Those bounds are for averages over timelike curves or worldvolumes. As an example, the renormalized null energy of the quantum massless minimally coupled scalar field in Minkowski spacetime averaged over a smooth timelike curve $\gamma$ obeys the following QEI\cite{Fewster:2002ne}
\begin{equation}
	\label{eqn:timelikeqei}
		\int dt \langle \nord{T_{\mu \nu}} \ell^\mu \ell^\nu \rangle_\omega f^2(t) \geq -\frac{(v^\mu \ell_\mu)}{12\pi^2} \int dt f''(t)^2  \,.
\end{equation}
for all Hadamard states $\omega$ and any smooth, real-valued compactly supported function $f$. Here $v^\mu$ is the timelike vector tangent to $\gamma$. For $f$ a normalized Gaussian with zero mean and $t_0$ variance the right hand side of \eqref{eqn:timelikeqei} becomes
\begin{equation}
		\int dt \langle \nord{T_{\mu \nu}} \ell^\mu \ell^\nu \rangle_\omega f^2(t) \geq -\frac{(v^\mu \ell_\mu)}{64\pi^2 t_0^4}  \,.
\end{equation}
Then we can see the physical interpretation of the QEI: the longer the timescale $t_0$, the less negative null energy is allowed. 

Important classical relativity results such as the Penrose singularity theorem\cite{Penrose:1964wq} have the NEC in their hypotheses. If one wants to apply such theorems in a semiclassical setting it is necessary to replace the pointwise energy condition with a condition obeyed by quantum fields, namely a QEI. As the Penrose theorem proves null geodesic incompleteness, the relevant QEI would be a null averaged one. 

The purpose of this contribution is to motivate a null QEI, the smeared null energy condition (SNEC) and use it to prove a semiclassical singularity theorem for null geodesic incompleteness. This theorem is applicable to the case of evaporating black holes. We begin with a description of the challenges to develop a null QEI and motivation for SNEC in Sec.~\ref{sec:null}. In Sec.~\ref{sec:singularity} we state the singularity theorem of Ref.~\citenum{Fewster:2019bjg} and show that SNEC can be used as an assumption. In Sec.~\ref{sec:blackhole} we apply the theorem to a toy model of evaporating black holes. We conclude in Sec.~\ref{sec:conclusions} with a summary and discussion of future work.

\section{Null quantum energy inequalities}
\label{sec:null}

\subsection{The Fewster-Roman counterexample}

In the expression of Eq.~\eqref{eqn:timelikeqei} the renormalized null energy is averaged over a timelike curve. A similar expression integrated over a null geodesic has been derived two-dimensions. In particular, Fewster and Hollands\cite{Fewster:2004nj} showed that
\begin{equation}
	\label{eqn:twod}
	\int^{+\infty}_{-\infty} f(\lambda) \langle \nord{T_{\mu \nu}} \ell^\mu \ell^\nu \rangle_\omega \geq - \frac{c}{48\pi}  \int^{+\infty}_{-\infty}  \frac{\left( f' \right)^2}{f} d\lambda \,,
\end{equation}
holds for a class of interacting quantum fields, namely the unitary, positive energy conformal field theories (CFTs) with stress-energy tensor in Minkowski spacetime. Here $c$ is the central charge of the theory. We recently generalized that result for a large class of curved backgrounds\cite{Freivogel:2020hiz}. 

The situation is different in more than two dimensions. Fewster and Roman\cite{Fewster:2002ne} showed using an explicit construction, that the renormalized null energy averaged over a null geodesic is unbounded from below for the massless minimally coupled scalar field. So there are no null QEIs in four-dimensional Minkowski space.

Their construction was a sequence of vacuum-plus-two-particle states. Then they allowed the three-momenta of excited modes to become increasingly parallel to the spatial part of the null vector $\ell^\mu$. As the three momenta grows, the lower bound of the inequality diverges to negative infinity. 

\subsection{The smeared null energy condition}

To overcome the problem encountered by Fewster and Roman, Freivogel and Krommydas proposed the smeared null energy condition (SNEC)\cite{Freivogel:2018gxj}. The main concept behind it is that in quantum filed theory there often exists an ultraviolet cutoff $\luv$. It was shown\cite{Freivogel:2018gxj,Freivogel:2020hiz} that the existence of a cutoff restricts the three momenta of the excited modes in the Fewster-Roman counterexample, leading to a finite lower bound. 

The SNEC can be written as 
\begin{equation}
\label{eqn:SNEC2}
\int^{+\infty}_{-\infty} d\lambda g^2(\lambda) \langle T_{kk} (x^{\mu}(\lambda)) \rangle \geq - \frac{4B}{G_N} \int^{+\infty}_{-\infty} d \lambda \left(g'(\lambda)\right)^2 \,.
\end{equation}
where $x^\mu(\lambda)$ is a null geodesic, $g(\lambda)$ is a differentiable `smearing function' that controls the region where the null energy is averaged, $B$ is a constant and $G_N$ is the Newton constant. In four-dimensional field theory we can write
\begin{equation}
\label{eqn:GNUV}
N G_N \lesssim \luv^2 \,,
\end{equation}
where $N$ is the number of fields. This relationship means the SNEC provides a finite lower bound even for a large number of fields. This is particularly useful for applications where the negative energy arises from multiple fields with small negative energy fluctuations (see e.g. \citenum{Maldacena:2018gjk}).

To have $B$ be an order one number, we need to saturate that inequality. This is the case for the induced gravity proof of \citenum{Leichenauer:2018tnq} where they derived $B=1/32\pi$. However, it is reasonable to consider a $B\ll 1$ since \eqref{eqn:GNUV} is typically not saturated in controlled constructions.

SNEC has been proven to hold for free fields on Minkowski spacetime \cite{Freivogel:2020hiz, Fliss:2021gdz}. The proof utilizes the fact that free field theory factorizes on the lightsheet in a collection of two-dimensional CFTs. For each of those CFTs the two-dimensional null QEI of Eq.~\eqref{eqn:twod} holds leading to a proof for the higher dimensional theory.

\subsection{The double smeared null energy condition}

It is unclear if the proof of SNEC described in the previous subsection can be generalized to curved spacetimes and interacting fields. Additionally, the bound diverges fro $\luv \to 0$, with the ultraviolet cutoff depending on the theory. 

Those disadvantages of SNEC led to the proposal of a different bound, the double smeared null energy condition (DSNEC). The idea is to average the renormalized null energy density in both null directions, denote $+$ and $-$. Schematically the DSNEC can be written as 
\begin{equation}
\int d^2x^\pm g^2(x^+,x^-)	\langle \nord{T_{--}}  \rangle_\omega \geq  -\frac{\mathcal N}{\delta_+ \delta_-^3} \,,
\end{equation}
where $\mathcal N$ depends on the number of fields and the smearing function, and $\delta_\pm$ is the smearing length in each of the null directions. 

For a massless scalar in Minkowski the DSNEC can be explicitly written as
\begin{equation}
\label{eqn:fourdposbound}
\int d^2x^\pm g^2(x^+,x^-)\langle \nord{T}_{--}\rangle_\omega \geq-A  \left[\int dx^+ (g_+''(x^+))^2\right]^{1/4} \left[\int dx^- (g_-''(x^-))^2\right]^{3/4} \,,
\end{equation}
where $A$ is a number and we assumed that the smearing function factorizes as $g^2=g_+(x^+)^2 g_-(x^-)^2$.

The DSNEC was motivated in \citenum{Fliss:2021gdz} and its rigorous proof will appear in future work\cite{Fliss:2021un}. The proof of DSNEC can straightforwardly generalized for curved spacetimes as it is derived from a general QEI valid in spacetimes with curvature\cite{Fewster:2007rh}. It includes no theory dependent cutoff and the smearing in each direction can be controlled. However, it is still unclear if the DSNEC can be used to prove singularity theorems. The main obstacle is that the usual proofs of those theorems require bounds on single geodesics.

\section{The singularity theorem}
\label{sec:singularity}

\subsection{The Penrose singularity theorem}

In general relativity a spacetime is singular if it possesses at least one incomplete and inextendible geodesic. This definition does not give us information about the nature of the singularity (e.g. if curvature scalars diverge) but it allowed for the first model-independent theorems, the singularity theorems of Penrose\cite{Penrose:1964wq} and Hawking \cite{Hawking:1966sx}.

Most singularity theorems have the same three types of hypotheses: the energy condition, the initial or boundary condition and the causality condition. In the case of the Penrose theorem the energy condition is the NEC or more accurately the geometric form of the NEC, the null convergence condition. The boundary condition is the existence of a trapped surface, a co-dimension two spacelike submanifold which has two null normals with negative expansions. Equivalently, a trapped surface has negative null normal curvature everywhere. Finally, the causality condition is the existence of a non-compact Cauchy hypersurface. The conclusion is that the spacetime is future null geodesically incomplete. 

Schematically, singularity theorems work in the following way: the initial condition establishes the convergence of a congruence of geodesics. The energy condition guarantees that the convergence will continue and a focal point will form. Finally, the causality condition does not allow the formation of focal points leading to a contradiction that proves the geodesic incompleteness.  

\subsection{Singularity theorems with weaker conditions}

As quantum fields violate all pointwise energy conditions, a semiclassical singularity theorem is required to have a weaker energy condition. Examples of singularity theorems with such conditions include Refs.~\citenum{Tipler:1978zz, ChiconeEhrlich:1980} and \citenum{Borde:1987qr} but none of them address the case of a condition obeyed by quantum fields. First Ref.~\citenum{Fewster:2010gm} proved singularity theorems with energy conditions inspired by QEIs. Ref.~\citenum{Fewster:2019bjg} proved singularity theorems with similar conditions using index form methods. Utilizing these results Ref.~\citenum{Fewster:2021mmz} proved the first semiclassical singularity theorem for timelike geodesic incompleteness. Here we follow Ref.~\citenum{Fewster:2021mmz} theorem for null geodesic incompleteness. 

To state the theorem we first need to fix a parametrization of the affine parameter of the null geodesic. For a manifold $M$ let $P$ be submanifold of co-dimension $2$ with mean normal curvature vector field $H^\mu=H \hat{H}^\mu$ where $\hat{H}^\mu$ is a future-pointing timelike unit vector. Then let $\gamma$ be a future-directed null geodesic emanating normally from $P$. Then $\hat{H}_\mu$ is extended by parallel transporting along $\gamma$. Now we can choose an affine parameter $\lambda$ on $\gamma$, such that $\hat{H}_\mu d\gamma^\mu/d\lambda =1$. Now we can state the energy condition from \citenum{Fewster:2019bjg}
\begin{equation}
\label{eqn:Riccinull}
\int_0^\ell g(\lambda)^2 R_{\mu \nu} \ell^\mu \ell^\nu d\lambda \geq -Q_m(\gamma) \|  g^{(m)} \|^2-Q_0 (\gamma) \| g\|^2 \,,
\end{equation}
where $Q_m$ and $Q_0$ are unknown constants dependent on the choice of $\gamma$ and $m$ a positive integer. The notation $\| \cdot \|$ denotes the $L^2$ norm.

The bound required by the singularity theorem is a geometric assumption, while SNEC is an assumption on the renormalized stress-energy tensor. Classically, the Einstein equation connects curvature to the stress-energy tensor. Semiclassically, the semiclassical Einstein equation (SEE) equates the expectation value of the stress-energy tensor with the classical Einstein tensor
\begin{equation}
\label{eqn:SEE}
8\pi G_N \langle T_{\mu \nu} \rangle_\omega= G_{\mu \nu} \,.
\end{equation}
Using the SEE the bound of eq.\eqref{eqn:SNEC2} can be written as
\begin{equation}
\label{eqn:firstbound}
\int_{-\infty}^\infty g(\lambda)^2 R_{k k} d\lambda \geq -32 \pi B \|g'(\lambda)\|^2 \,.
\end{equation}
Then this is a bound of the form of eq.\eqref{eqn:Riccinull} with $m=1$, $Q_1=32 \pi B$ and $Q_0=0$. Using the SEE assumes that we have a self-consistent solution, which includes a state $\omega$ and a metric $g_{\mu \nu}$.

In addition to the energy condition the theorem of Ref.~\citenum{Fewster:2019bjg} has an assumption on the pointwise null energy density for a finite affine parameter. In particular, there are two scenarios to describe all possible initial conditions: in scenario 1, initially the NEC is satisfied for an affine length $\ell_0$, short compared to the one for the formation of a focal point $\ell$. In scenario 2 this requirement is dropped and instead conditions are imposed on the null contracted Ricci tensor for small negative values of the affine parameter. Here we focus on scenario 2. 

We first extend $\gamma$ to $\gamma:[-\ell_0,\ell]\to M$ and assume that eq.\eqref{eqn:firstbound} holds on the extended geodesic. Then we define $\rho_{\max} = \max_{[-\ell_0,0]} \rho$ and we can use Lemma 4.7 of Ref.~\citenum{Fewster:2019bjg} with $m=1$, $Q_0=0$, $A_1=1/3$, $B_1=C_1=1$. If we additionally assume that $\rho_{\max}<0$  we have

\begin{lemma}\label{lem:scenario2null}
	For $\rho$ satisfying eq.\eqref{eqn:firstbound} on $[-\ell_0,\ell]$ if 
	\begin{equation}
		\label{eqn:Hnull2}
		-2H \geq \frac{Q_1+2}{\ell}+\frac{Q_1}{\ell_0}+\frac{1}{3} \rho_{\max} \ell_0 \,.
	\end{equation}
	then there is a focal point to $P$ along $\gamma$ in $[0,\ell]$.  
\end{lemma} 
Negative null energy in $[-\ell_0,0]$ region leads to smaller required initial contraction because this negative energy must be over-compensated by positive energy. This effect has been studied and it is known as ``quantum interest''\cite{Ford:1999qv}.

\section{Application to evaporating black holes}
\label{sec:blackhole}

Penrose proved the first singularity theorem which applies to a classical black hole spacetime. However, this theorem cannot be applied in an evaporating black hole spacetime, where  the NEC is violated. Here we apply Lemma~\ref{lem:scenario2null} in a toy model of an evaporating black hole spacetime.

First, we assume that the metric is approximated by Schwarzschild geometry near the classical horizon
\begin{equation}
ds^2=\left(\frac{R_s}{r}-1 \right) dt^2-\left(\frac{R_s}{r}-1 \right)^{-1} dr^2+r^2 d\Omega^2 \,,
\end{equation}
where $R_s$ is the Schwarzschild radius. We focus on spherically symmetric hypersurfaces $P$, so that the hypersurface is defined by Schwarzschild coordinates $(t_p, r_p)$ where the mean normal curvature vector field is purely in the $r$ direction. 

Inside the horizon, the mean normal curvature $H$ of our surfaces $P$ is given by\cite{Pilkington:2011aj}
\begin{equation}
\label{eqn:Hsch}
H(r_P)=-\frac{1}{r_P} \sqrt{\frac{R_s}{r_P}-1} \,.
\end{equation}

Since we assumed $ \rho_{\max}<0$ we can drop the last term of Eq.\eqref{eqn:Hnull2}
\begin{equation}
\label{eqn:Happl}
H< -\frac{Q_1}{2\ell}-\frac{1}{\ell} -\frac{Q_1}{2\ell_0} \,.
\end{equation}
$H$ depends on two parameters, the maximum affine parameter for the formation of the singularity $\ell$ and the length of the affine parameter that the NEC is violated $\ell_0$.

We define the dimensionless parameter $x$ 
\begin{equation}
R_s - r_P \equiv x R_s \,, \qquad 0< x <1 ~,
\end{equation}
and $y$ by demanding that the affine distance $\ell$ is a coordinate distance $yR_s$. We can consider the case that $y \to \infty$ meaning we have no information about the location of the singularity. 

\begin{figure}
	\centering
	\includegraphics[height=5cm]{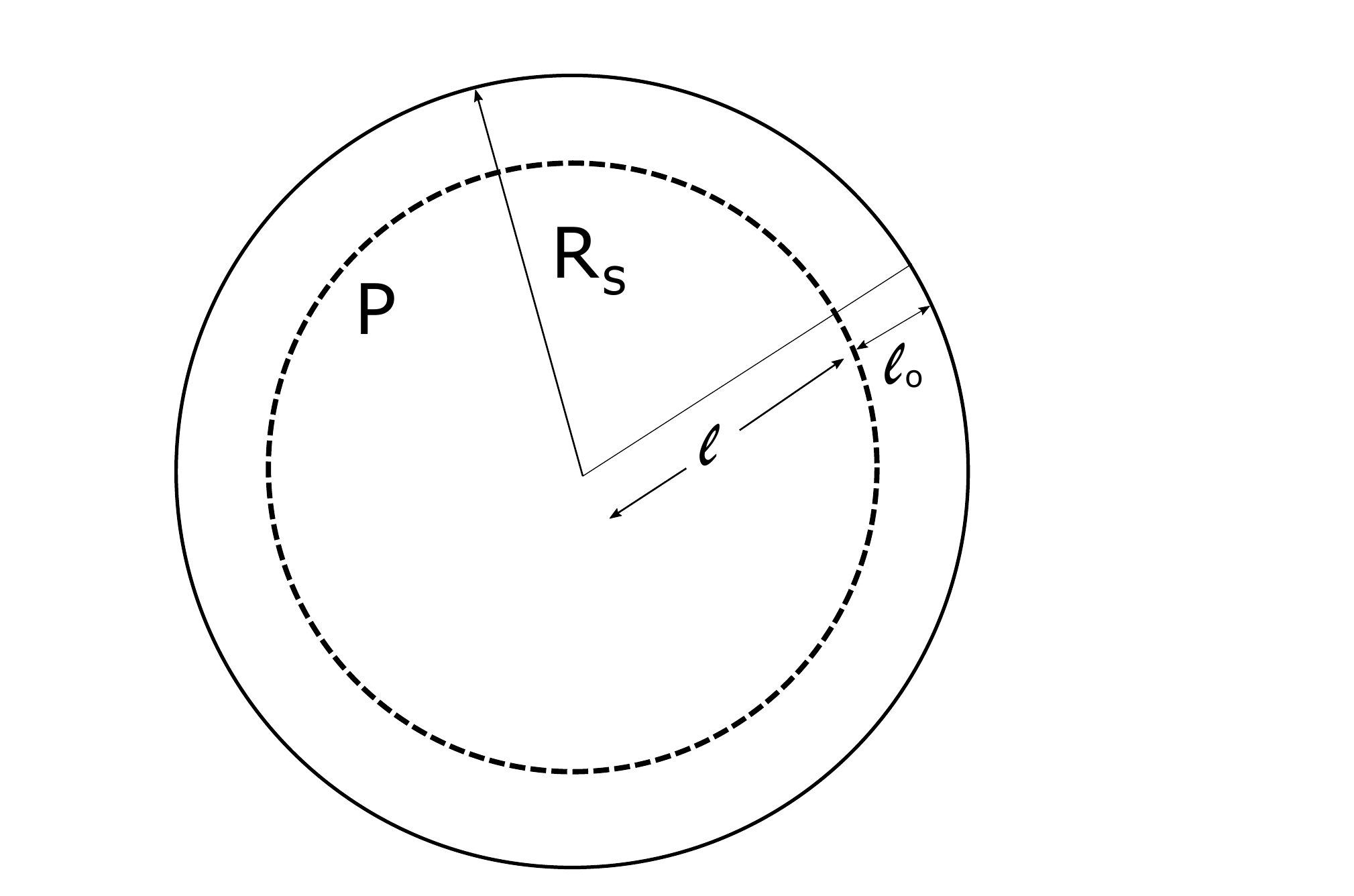}
	\caption{Schematic representation of a Schwarzschild black hole and the parameters. The dashed circle is constant $r$ and $t$ hypersurface $P$. Distance $\ell_0$ is from the point where the NEC starts being violated, and distance $\ell$ is from $P$ to the singularity (pictured here at $r=0$).}
	\label{fig:bh}
\end{figure}

The idea is that if the mean normal curvature of the hypersurfaces $P$ is smaller than the one required by Lemma~\ref{lem:scenario2null} we have a singularity. So we equate the expressions \eqref{eqn:Hsch} and \eqref{eqn:Happl} to find the location of the first hypersurface inside the horizon for which we can apply the Lemma. We want $P$ to be as close to the classical horizon as possible. The setup is shown in Fig.~\ref{fig:bh}.

\begin{figure}
	\centering
	\includegraphics[height=6cm]{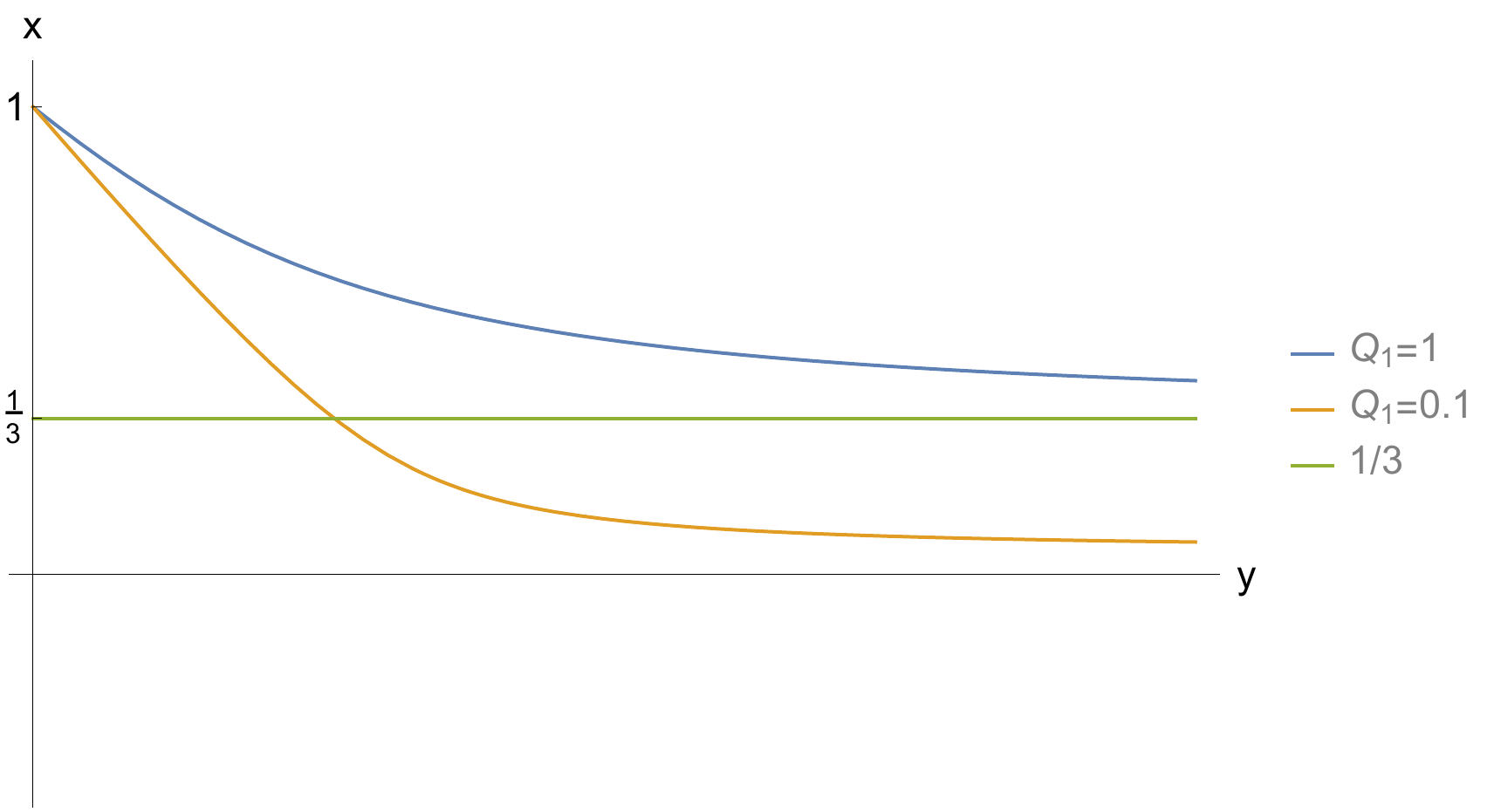}
	\caption{Required value of $x$ to have a singularity for different values $y$. For $Q_1=0.1$ the minimum value is much smaller compared to $Q_1=1$.}
	\label{fig:scen2}
\end{figure}

 A plot of  $x$ for different values of $y$ is shown in Fig.~\ref{fig:scen2} for two different values of $Q_1$. The Ref.~\citenum{Leichenauer:2018tnq} value of $B=1/32\pi$ translates to $Q_1=1$. Using this value for $Q_1$, we find that the minimum $x$ is $1/3$. As discussed earlier, there is also strong motivation to use a value of $B\ll 1$ and so $Q_1\ll1$. For small $Q_1$, we have a singularity theorem for spheres $P$ with
 \begin{equation}
 R_s - r_P \gtrapprox R_s \frac{Q_1}{2} \ \ \ \ \ \ {\rm  for}\ Q_1 \ll 1~.
 \end{equation}

\section{Conclusions}
\label{sec:conclusions}

In this work we provided motivation for both the smeared null energy condition (SNEC) and the double smeared null energy condition (DSNEC). We proved a semiclassical singularity theorem using SNEC and applied this theorem to establish that spacetimes that approximate the Schwarzschild solution near the horizon must contain a singularity.

As the version of SNEC we use for the singularity theorem has only been proven for Minkowski spacetime, an important future direction is a theorem with a condition that incorporates curvature. There are different ways to approach that. One is to attempt to prove SNEC for spacetimes with curvature. Another is to use the new bound of DSNEC instead. This would require novel concepts as the current singularity theorem proofs are for bounds on single geodesics.

\section*{Acknowledgments}
BF and E-AK are supported by the ERC Consolidator Grant QUANTIVIOL. This work is part of the $\Delta$ ITP consortium, a program of the NWO that is funded by the Dutch Ministry of Education, Culture and Science (OCW).

\bibliographystyle{ws-procs961x669}
\bibliography{ws-pro-sample}

\end{document}